\documentclass[10pt,aps,prl,twocolumn,superscriptaddress,floatfix,showpacs,longbibliography]{revtex4-1}

\usepackage{graphicx}
\graphicspath{{./}{./states/}}

\usepackage{dcolumn}
\usepackage{bm}
\usepackage{hyperref}

\usepackage[usenames,dvipsnames,table]{xcolor}
\usepackage[version=3]{mhchem} 
\usepackage{braket}
\usepackage{multirow}
\usepackage{soul}

\usepackage{ulem} 

\hypersetup{hidelinks}
\sethlcolor{mark}

\begin{document}

\title{Probing the topology of the quantum analog of a classical skyrmion}

\author{O. M. Sotnikov}
\affiliation{Theoretical Physics and Applied Mathematics Department, Ural Federal University, Mira Str. 19, 620002 Ekaterinburg, Russia}

\author{V. V. Mazurenko}
\email{vmazurenko2011@gmail.com}
\affiliation{Theoretical Physics and Applied Mathematics Department, Ural Federal University, Mira Str. 19, 620002 Ekaterinburg, Russia}

\author{J. Colbois}
\affiliation{Institute of Physics, \'Ecole Polytechnique F\'ed\'erale de Lausanne (EPFL), CH-1015 Lausanne, Switzerland}

\author{F. Mila}
\affiliation{Institute of Physics, \'Ecole Polytechnique F\'ed\'erale de Lausanne (EPFL), CH-1015 Lausanne, Switzerland}

\author{M. I. Katsnelson}
\affiliation{Radboud University, Institute for Molecules and Materials, Nijmegen, Netherlands}
\affiliation{Theoretical Physics and Applied Mathematics Department, Ural Federal University, Mira Str. 19, 620002 Ekaterinburg, Russia}

\author{E. A. Stepanov}	
\affiliation{I. Institute of Theoretical Physics, University of Hamburg, Jungiusstrasse 9, 20355 Hamburg, Germany}
\affiliation{Theoretical Physics and Applied Mathematics Department, Ural Federal University, Mira Str. 19, 620002 Ekaterinburg, Russia}

\begin{abstract}
In magnetism, skyrmions correspond to classical three-dimensional spin textures characterized by a topological invariant that keeps track of the winding of the magnetization in real space, a property that cannot be easily generalized to the quantum case since the orientation of a quantum spin is in general ill-defined. 
Moreover, as we show, the quantum skyrmion state cannot be directly observed in modern experiments that probe the local magnetization of the system. 
However, we show that this novel quantum state can still be identified and fully characterized by a special local three-spin correlation function defined on neighbouring lattice sites -- the  scalar chirality -- which reduces to the classical topological invariant for large systems, and which is shown to be nearly constant in the quantum skyrmion phase.
\end{abstract}

\maketitle	

The broad use of topological language is one of the main trends in contemporary physics, including condensed matter physics and even materials science \cite{Schap89,Thou,Naka,Volo,Wen1,Wen2,Kats20,Merm,Qi10,Hald,Kost}. Numerous nontrivial topological effects in superfluid helium-3~\cite{Volo}, the concept of topological quantum phases in strongly correlated systems~\cite{Wen1,Wen2}, topologically protected zero-energy states in magnetic field and other topology-related issues in graphene~\cite{Kats20}, and the quickly growing field of topological insulators~\cite{Qi10} provide excellent examples. When considering quantum systems we usually deal with topology in reciprocal $k$-space~\cite{Wen1,Wen2}, whereas for classical systems topological protection of defects of different kind ~\cite{Merm} plays a crucial role. 

Among such defects, magnetic skyrmions~\cite{bogdanov1989thermodynamically} are currently attracting special attention due to perspectives to use them in magnetic information storage~\cite{Jonietz1648, fert2013skyrmions, Tomasello:2014qd}. The progress in the development of experimental techniques~\cite{Muhlbauer915, PhysRevLett.102.186602, PhysRevB.81.041203, yu2010real, nagaosa2013topological, TsirlinCu,X-ray,X-ray1} poses new challenges for the theory and numerical simulations of ordered magnetic phases~\cite{roadmap2020}. 
%
Thus, skyrmions with the characteristic size of a few nanometers have already been observed experimentally~\cite{Wiesendanger, Romming636}, and predicted theoretically in frustrated magnets~\cite{PhysRevLett.108.017206, leonov2015multiply, PhysRevB.93.064430}, narrow band Mott insulators under high-frequency light irradiation~\cite{PhysRevLett.118.157201}, and Heisenberg-exchange-free systems~\cite{skyrm_classic}. On such small characteristic length scales compared to the lattice constant, quantum effects cannot be neglected. 
The same difficulty also arises in low-dimensional systems with small spin (e.g. $S=1/2$)~\cite{Schliemann} and itinerant systems with delocalized magnetic moments.

It is however not at all clear what a quantum skyrmion (that is, a skyrmion in  a system of quantum spins) could be. There is indeed no way to introduce a topological charge for the {\it quantum} spin case which would protect quantum skyrmions similar to the topological protection in classical systems~\cite{Merm}. Physically it means that topologically protected classical spin configurations are, generally speaking, not robust with respect to quantum tunneling which can transform them into topologically trivial states. Nevertheless one can assume that the existence of topologically protected {\it classical} magnetic configurations should influence, in some way, the properties of the corresponding quantum systems.


This fundamental problem was
not clarified in the previous attempts to introduce ``quantum skyrmions''.
Instead, description of this quantum state was
either done semiclassically assuming that the magnetization
dynamics
is dominated by classical magnetic excitations that emerge on top of the symmetry-broken ground state of the system~\cite{PhysRevB.94.134415, PhysRevLett.124.157203, 2018arXiv180702203O}, or by means of the Holstein-Primakoff transformation, which only allows one to compute quantum corrections to the classical solution~\cite{PhysRevB.92.245436}.
Also, the standard identification of a skyrmion by its magnetization pattern was used in recent works~\cite{Rosch, Lorente}, where topological states of small clusters embedded in a ferromagnetic environment were investigated. 
However, this does not take into account the fact that,
strictly speaking, the corresponding states are not actually ``topological'' in the sense of some rigorous protection. This problem will be addressed in this work.

\begin{figure}[!t]
\includegraphics[width=0.95\columnwidth]{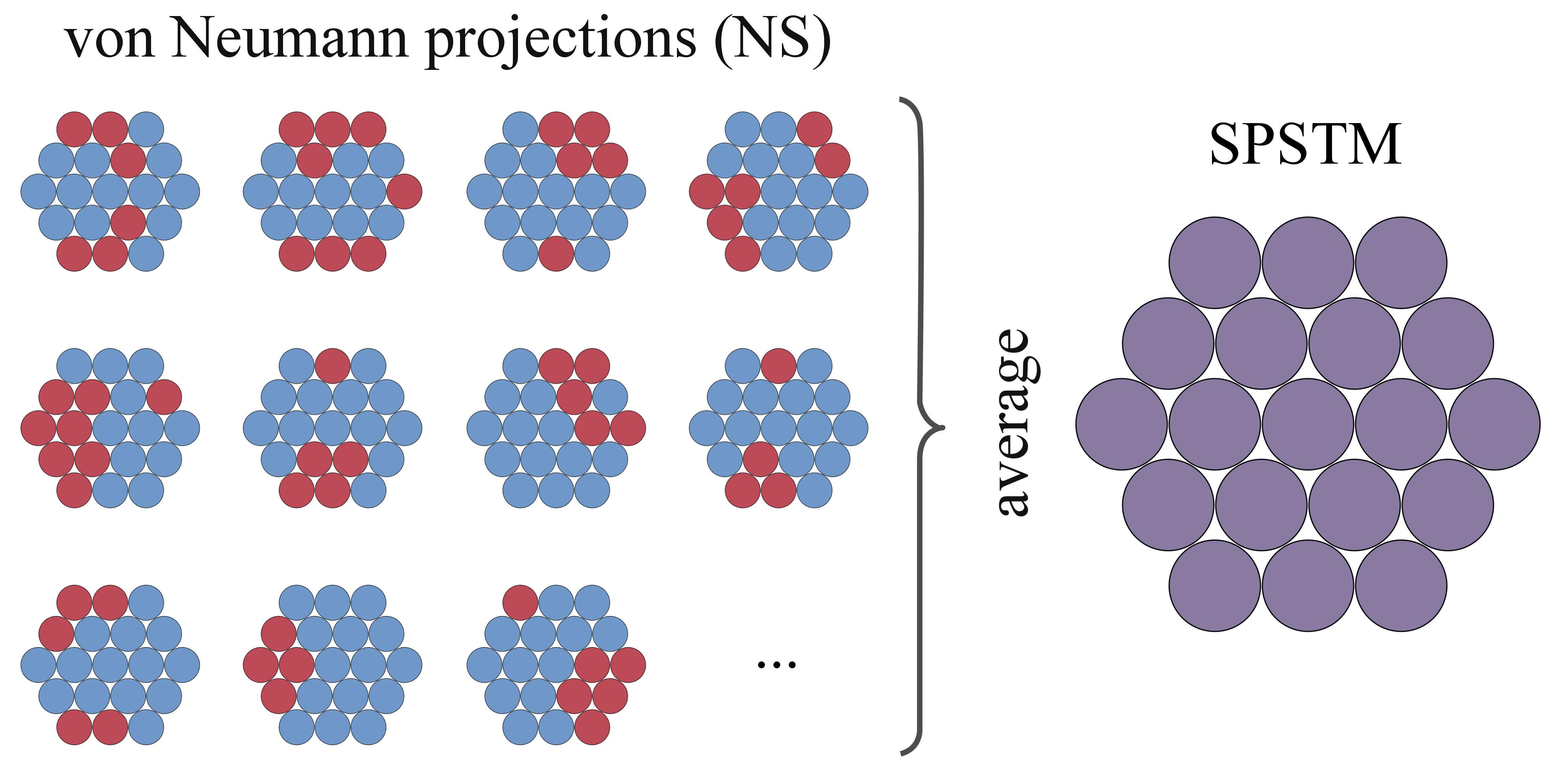}
\caption{\label{fig:start}
Schematic representation of the local magnetization measurements of the quantum skyrmion state realized on a 19-site cluster.
Upon individual von Neumann measurements, this state collapses to different basis functions shown in left panel. The average of the local magnetization over all basis functions results in a uniform magnetization pattern that could be observed in SPSTM experiment (right panel). }
\end{figure}

{\it Characterization of a quantum skyrmion} ---
Conceptually, the quantum skyrmion problem is somewhat similar to the formation of the antiferromagnetic (AFM) ordering in quantum systems.
Whereas the classical skyrmion solution on the lattice is characterized by a distinct magnetic pattern, in the infinite quantum system all lattice sites are identical and have the same value of the local magnetization. 
Assuming that modern Lorentz and spin-polarized scanning tunneling microscopy (SPSTM) techniques weakly affect the quantum system, the measurement of the quantum skyrmion state will thus result in the same value of the local magnetization for all lattice sites (see Fig.~\ref{fig:start} right). 
In the case of antiferromagnets, the appearance of sublattices is expected to be induced by applying a staggered field that selects the classical N\'eel state from the quantum solution of the problem, but,  contrary to the case of ferromagnets, this field is not very physical, and it is not easy at all to understand how it can be realized in practice. However, the same can be achieved with the use of the analog of the quantum Zeno effect~\cite{doi:10.1063/1.523304, joos2013decoherence}, that is, by performing repeated local (von Neumann) measurements of the spin~\cite{Hafer,AFM3, AFM2}, which can possibly be realized in inelastic X-ray or neutron scattering (NS) experiments. This strong influence on the quantum system will result in the collapse of the wave function to one of the possible basis states with a certain probability, via formation of the ``decoherence waves''~\cite{AFM3}.

For the quantum skyrmion case, it turns out that these states do  not resemble the classical skyrmion solution.
To demonstrate this point, we consider a particular example, the quantum skyrmion state discussed below~\eqref{eq:ham} stabilized on a 19-site cluster with periodic boundary conditions at magnetic field $B=0.4$.
Von Neumann measurements of the local magnetization are modelled using a quantum computer simulator as implemented in \texttt{QISKIT} package~\cite{Qiskit}. To this aim we obtain the ground state of the considered system and use it for initialization of qubits.
After that, each qubit is measured several times to get different basis functions demonstrated in left panel of Fig.~\ref{fig:start}.
Since the considered quantum system is translationally invariant, the local magnetization of the ground state averaged over all basis functions $\braket{ \hat S^z_i} = \braket{\Psi_0 | \hat S^z_i | \Psi_0}$ is uniform.
Therefore, contrary to the classical skyrmion case, the quantum skyrmion state cannot be detected in any modern experiment that performs a simple local measurement of the magnetization.

Instead, one could calculate the momentum-space representation of the more complicated spin-spin correlation function (structure factor), and if there is magnetic ordering, it is signalled by the development of Bragg peaks at momenta that correspond to the wave vectors ${\bf q}$ of the ordering, as in antiferromagnets.
Since the classical skyrmion can be considered as a superposition of spin spirals, a similar pattern of Bragg peaks is expected for a quantum skyrmion state.
However, due to the quantum nature of the problem,  and by contrast to the classical case, the quantum helical phase is also characterized by the quantum superposition of spin spirals.
Therefore, as we demonstrate below, the structure factor also does not allow to distinguish between these two phases of the quantum system.


Strictly speaking, even in the classical case the spin-spin correlation function is also not a sufficient measure for a skyrmion state, because different skyrmion, vortex, bubble and multi-domain phases are indistinguishable on the level of the structure factor (see e.g.~\cite{Maz1, Maz2}), and some more complicated correlation functions should be used as discussed in the Supplemental Material (SM)~\cite{suppl}.
The classical skyrmion state is actually characterized by a topological invariant which, for continuum models of magnetism, is given by the following expression
\begin{align}
Q = \frac{1}{4\pi}\int {\bf m} \cdot \left[ \partial_{x}{\bf m} \times \partial_{y}{\bf m} \right] dx \, dy
\label{eq:Q_class}
\end{align}
that counts the number of times the magnetization ${\bf m}(r)$ wraps around a sphere. This characterization depends in an essential way on the relative orientation of the local spins, an information which cannot be extracted from the quantum ground state for the reasons explained above.

The fundamental problem is thus how to generalize the classical topological invariant~\eqref{eq:Q_class} to the quantum case. On a lattice, the proper version of the classical topological invariant has been proposed by Berg and L\"uscher~\cite{luscher}.
According to their idea, the winding of the magnetization can be approximated by a sum of all spherical surfaces that are formed by three neighboring spins. 
Unfortunately, as shown in SM~\cite{suppl}, their expression for the skyrmion number $Q_{BL}$ cannot be easily converted into a linear quantum operator. 
What we propose here is to use a discrete version of the topological invariant, the scalar chirality (see below). 
As we shall see, this quantity, which is naturally defined for both classical and quantum spins, captures the non-collinearity of neighboring spins, and it turns out to be almost constant inside skyrmion phases both for classical and quantum spins.
In the quantum case, the scalar chirality reduces to a local quantity defined for nearest-neighbor spin operators leading to a general and flexible characterization of skyrmions. 

\begin{figure*}[t!]
\includegraphics[width=0.9\linewidth]{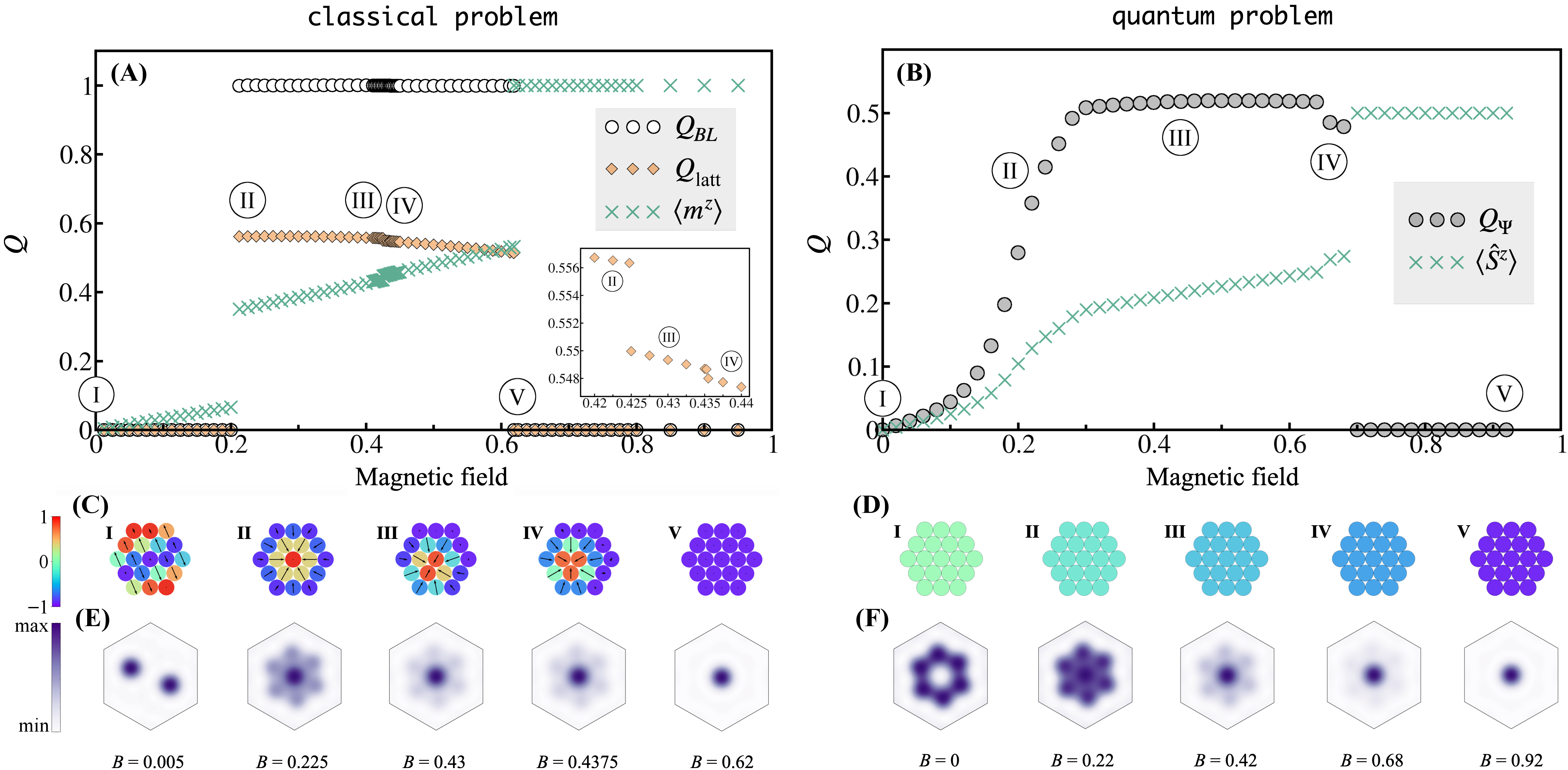}
\caption {\label{fig:summary} Complete set of observables describing skyrmions in the classical and quantum case. Skyrmion number and average magnetization (A, B), local magnetization pattern of the lattice (C, D), and structure factors (E, F) as a function of magnetic field for the classical (left panels) and quantum (right panels) problems for a 19-site triangular lattice with periodic boundary conditions. 
Roman numbers denote different phases. Since the classical ground state is many-fold degenerate we have chosen a representative pattern for the magnetization panel (C). The inset in panel (A) shows three different types of classical skyrmions revealed in the intermediate phase by the scalar chirality. }
\end{figure*}

{\it Results} --- We start with the following lattice Hamiltonian of a quantum spin model
\begin{align}
\hat H = \sum_{ij} J_{ij} \hat{\bf S}_i  \cdot  \hat{\bf S}_j + \sum_{ij} {\bf D}_{ij} [\hat{\bf S}_i \times \hat{\bf S}_j] + \sum_{i} B^z \hat{S}^z_i.
\label{eq:ham}
\end{align}
Here, $J_{ij}$ is the isotropic Heisenberg exchange interaction. $\mathbf{D}_{ij}$ is an in-plane vector that points in the direction perpendicular to the bond between neighboring $i$ and $j$ sites and describes the Dzyaloshinskii-Moriya interaction (DMI). $\mathbf{B}$ is an external uniform magnetic field applied along the $z$ direction. 
Quite generally, the competition between the exchange interaction and the DMI leads to the formation of a classical skyrmion that is usually stabilized by a nonzero magnetic field.

Let us look 
at the phase diagram of the model of Eq.~(\ref{eq:ham}) on the triangular lattice. To compare the classical and the quantum spin-1/2 cases, we have chosen to work on a 19-site cluster that is one of the largest systems for which the ED solution can be obtained~\cite{CRSMatrix}. The exchange interaction is set to $J=-0.5 D$, where $D$ is the length of the DMI vectors, to produce a nanoskyrmion compatible with this cluster size. The main results are summarized in Fig.~\ref{fig:summary}. The classical model has been solved for unit magnetization vectors. After that classical energies calculated for different magnetic fields have been normalized to compare the result with the quantum solution.

For the classical case, the presence of three main phases is already clear from the average magnetization $\langle m^{z} \rangle$ (Fig.~\ref{fig:summary}\,(A)), which exhibits two jumps at $B_{c1}\simeq 0.21$ and $B_{c2}\simeq 0.6$,  indicating strongly first order phase transitions and major reorientations of the spins. However, a closer look at the real space orientation of the spins (Fig.~\ref{fig:summary}\,(C)) shows that in the low-field phase the spins are coplanar (I), while in the intermediate phase the spins form three different 3d-textures (II, III, and IV) in spin space, which are hardly distinguishable from the energy plot (see the SM~\cite{suppl}). Above $B_{c2}\simeq 0.6$, the spins are fully polarized (V), so $B_{c2}$ is the saturation field. To further identify the nature of the various phases, it is useful to look at several additional properties. The first one is the static longitudinal spin structure factor defined as
$
X^{\parallel}_{\bf q} = \langle \hat S^{z}_{\bf q} \hat S^{z}_{-{\bf q}} \rangle.
$
Fig.~\ref{fig:summary}\,(E) shows that in the low field phase, it exhibits two Bragg peaks at opposite wave vectors ${\bf q}$ and $-{\bf q}$, typical of a helical state of pitch vector ${\bf q}$. In the intermediate phase, the structure factor is less specific. It looks like the superposition of 3 pairs of Bragg peaks rotated by $\pi /3$ and of a Bragg peak at the zone center. This is, of course, consistent with a skyrmion structure that, together with the superposition of enclosed spin spirals, is associated with the ferromagnetic ordering along the skyrmion boundary. However, as we have pointed out above, the real identification comes from the topological invariant $Q_{BL}$, which is calculated here using the Berg-L\"uscher approach~\cite{luscher}. As expected, we observe that this invariant is equal to unity in the intermediate phase and vanishes outside it (Fig.~\ref{fig:summary}\,(A)). Importantly, neither the structure factor nor the topological invariant can reflect the presence of three types of skyrmions in the intermediate phase.

Let us now try to perform a similar analysis for the quantum case. From the average magnetization (Fig.~\ref{fig:summary}\,(B)), three regimes emerge, but as compared to the classical case, the first transition is rounded.  
Indeed, $\langle \hat{S}^{z} \rangle$ shows a rapid but smooth increase at a field $B_{\Psi1}\simeq 0.3$ and a jump at $B_{\Psi2}\simeq 0.7$. The identification of the nature of these phases is by far not as simple however. First of all, as anticipated in the introduction, the expectation value of the local spin is uniform. So it is impossible to detect a planar or a $3d$-texture from this observable as can be seen from Fig.~\ref{fig:summary}\,(D). The natural idea is then to turn to the structure factor (Fig.~\ref{fig:summary}\,(F)). However, there is no qualitative difference between low and intermediate fields: in both cases, there are six maxima forming a hexagon and a maximum at the zone center, as in the skyrmion phase of the classical case. 

Does it mean that there is a single phase between zero field and saturation, and no well defined skyrmion phase? Not necessarily. Indeed, if we think in semiclassical terms, the effect of quantum fluctuations on a helical phase will be to stabilize a linear combination of helices if there are different choices of equivalent wave-vectors, and indeed here there are three equivalent choices of pitch vector. So if the low field phase is the quantum version of the helical phase, we indeed expect to have a hexagon of peaks. The problem is that this is also expected in the case of a skyrmion phase. So it is possible that there are two different phases for quantum spins as well. It is just impossible to distinguish them with the structure factor.

This example clearly calls for an alternative characterization of quantum skyrmions. The solution we propose is based on the following remarks. First of all, the fundamental difference between a classical helical state and a classical skyrmion is that the helical state is a coplanar structure (all spins lie in a given plane) while a skyrmion is a $3d$-texture. So these structures can be distinguished by the mixed product of three spins ${\bf S}_{i} \cdot [ {\bf S}_{j} \times {\bf S}_{k}]$, where $i$, $j$, and $k$ are three arbitrary lattice sites. Indeed this expression will be exactly zero for a helical state, but not for a skyrmion. In fact, the skyrmion invariant involves a similar mixed product of three magnetization vectors, because the discrete form of the classical topological invariant~\eqref{eq:Q_class} can be written as~\cite{Rosales}  
\begin{align}
Q_\text{latt} = \frac{1}{8\pi} \sum_{\langle{ijk\rangle}} {\bf m}_{i} \cdot [ {\bf m}_{j} \times {\bf m}_{k}]. 
\label{eq:Qlatt}
\end{align}
Here, ${\bf m}_{i}$, ${\bf m}_{j}$, and ${\bf m}_{k}$ are classical magnetization vectors of length $1$, and the summation runs over all non-equivalent elementary triangles that connect neighboring $i$, $j$, and $k$ sites. 
This quantity is known in other contexts as the {\it scalar chirality}, the term we will use from now on.

Importantly, as we show in SM~\cite{suppl}, the scalar chirality~\eqref{eq:Qlatt} coincides with the topological invariant~\eqref{eq:Q_class} in the classical limit of the skyrmion when the magnetization slowly varies with respect to a lattice constant.  
For nanoskyrmions, whose typical length scale is comparable to the lattice constant, a more precise result for the topological invariant is given by the Berg-L\"uscher approximation. Still, the scalar chirality is equally good when it comes to distinguishing a helical phase from a skyrmion phase. Indeed, it vanishes identically in a helical phase because it is strictly coplanar, and it does not for a 3d-texture. For the classical case this fact is illustrated in Fig.~\ref{fig:summary}\,(A).
A closer look at the scalar chirality in the intermediate range of magnetic fields presented in the inset allows one to distinguish three skyrmion phases. They are characterized by the different size and structure of the magnetic pattern, which is illustrated in Fig.~\ref{fig:summary}\,(C) on II, III, and IV panels. 
Thus, contrary to the topological number $Q_{BL}$, the scalar chirality is sensitive to different types of magnetic skyrmions.

Now, the main advantage of the scalar chirality over the Berg-L\"uscher invariant when it comes to quantum systems is that this quantity can be interpreted as a linear operator for quantum spins, so that a ground state indicator can be defined by simply calculating the expectation value of this operator in the ground state. This leads to the following simple definition of the quantum scalar chirality
\begin{align}
Q_{\Psi} = \frac{N}{\pi} \langle \hat {\bf S}_{1} \cdot [\hat {\bf S}_{2} \times \hat {\bf S}_{3}]  \rangle,
\label{QPsi}
\end{align}
where $N$ is the number of non-overlapping elementary triangular plaquettes that cover the lattice. Labels $1$, $2$, and $3$ depict three different spins that form an elementary plaquette. Here, we used the fact that the quantum ground state of the system is translationally invariant, so that the value of the scalar chirality is the same for any elementary triangle.
Therefore, the local three-spin correlation function defined on neighboring lattice sites~\eqref{QPsi} already gives complete information about the topology of the entire quantum system, something that is impossible in the classical case. 

As shown in Fig~\ref{fig:summary}\,(B), $Q_{\Psi}$ behaves differently in low, intermediate, and high-field phases. Contrary to the classical case, at low fields ($B_{\Psi1} < 0.3$) the quantum chirality increases gradually with the magnetization. Approaching the intermediate regime, $Q_{\Psi}$ saturates and remains nearly constant in a very broad range of magnetic fields. This remarkable result cannot be simply interpreted as a freezing of the system since the magnetization keeps growing as in the low field phase, implying that the quantum ground state of the system evolves continuously. Finally, at the critical field $B_{\Psi2}\simeq 0.7$, the system enters the fully polarized regime, as indicated by the stepwise decrease of the quantum chirality to zero. 
The physical picture for the low-field phase is that the ground state is coplanar at zero magnetic field. But instead of remaining coplanar as in the classical case, the linear combination of helical states in the quantum system acquires a non-coplanar structure upon increasing the field. In this case, spins progressively move out-of the plane in the direction of the field, which results in a nonzero value of the scalar chirality proportional to the tilting angle. 
By contrast, in the intermediate phase, the relative angle between spins does not change. It is the collective orientation of the skyrmion spin texture that allows the system to continue developing magnetization. 

{\it Conclusion} --- We have introduced and analyzed a novel quantum state of a spin system -- a quantum skyrmion. We have shown that this state can only be fully characterized by the expectation value of a skyrmion operator related to the local quantum scalar chirality of three neighboring spins. Indeed, in close analogy to the topological invariant that keeps track of the winding in classical skyrmions, the expectation value of the skyrmion operator is field independent to very high accuracy inside the skyrmion phase, by contrast to the simple superposition of spin orderings, where it changes a lot with the field. The value at which it stabilizes is related to the size of the skyrmion, and it would approach unity for very large skyrmions. This reduction factor is related to the value of the nearest-neighbor correlation function and can be independently estimated, so that, if necessary, the expectation value of the skyrmion operator could also be used to estimate the number of skyrmions in a quantum nano-skyrmion structure. 
We believe that our results can stimulate the development of experimental techniques to locally probe the topology of the entire quantum system.
For instance, the impact of the scalar chirality can be seen in the topological Hall effect~\cite{THE1, THE2, THE3, THE4},  in the finite topological orbital moment~\cite{TOM1, TOM2, TOM3, TOM4, TOM5, TOM6}, or in nonlinear optical experiments~\cite{SHG1, SHG2, SHG3, SHG4}.\\

\begin{acknowledgments}
We thank Sergey Brener for interesting discussions. The work of V.V.M., O.M.S., and E.A.S. was supported by the Russian Science Foundation Grant 18-12-00185. The work of J.C. and F.M. is supported by the Swiss National Science Foundation. The work of M.I.K. is supported by European Research Council via Synergy Grant 854843 - FASTCORR.
\end{acknowledgments}

\section{Supplementary Materials}

\section*{Method}

The eigenstates used for the calculation of the scalar chirality, magnetization and structure factors were obtained via an exact diagonalization approach. For that purpose we used the implicitly restarted Arnoldi algorithm as implemented in ARPACK library. Such a solution allowed us to optimize memory and CPU utilization due to CRS (Compressed Row Storage) sparse matrix format~\cite{CRSMatrix} used for the representation of the Hamiltonian. It is noticeable that, for the triangular supercell we have used, the ground state is sixfold degenerate for magnetic field values below 0.28 and in a narrow region between the skyrmion and saturated phases. To obtain such a rich ground state structure we calculated 24 extreme eigenstates with lowest energy using 256 Arnoldi vectors. We also checked that further increasing these numbers does not change the ground state structure. The calculation of scalar chirality as well as fidelity, local magnetization and structure factors was performed on GPU using CUDA framework and cuBLAS library.


\section*{Connection to a classical skyrmion}

In this section we show that the scalar chirality, which is a discrete form of the classical skyrmion number, coincides with the expression for the topological number proposed by Berg and L\"uscher \cite{luscher} in the continuous limit.
As has been mentioned in the main text, Berg and L\"uscher proposed to calculate the skyrmion number through the sum of all spherical surfaces that are formed by three neighboring $\langle ijk \rangle$ spins. The most convenient expression for the Berg and L\"uscher formula for the skyrmion number has been introduced in Ref.~\cite{Rosch}
\begin{align*}
Q_{BL} = \frac{1}{2\pi} \hspace{-1pt} \sum_{\langle ijk \rangle} \hspace{-1pt} \tan^{-1} \hspace{-2pt} \left[ \frac{8 \langle \hat{\bf S}_{i} \cdot [ \hat{\bf S}_{j} \times \hat{\bf S}_{k}]\rangle }{1 + 4 (\langle \hat{\bf S}_{i} \hat{\bf S}_{j}\rangle + \langle \hat{\bf S}_{j} \hat{\bf S}_{k}\rangle + \langle \hat{\bf S}_{k} \hat{\bf S}_{i}\rangle)} \right]
\end{align*}
where the authors replaced all spin products by correlation functions as an attempt to introduce a quantum analog of the classical skyrmion number. 
Since the quantum ground state is translationally invariant, all lattice sites are identical, and the neighboring scalar product and mixed product correlation functions on the triangular lattice are the same. This allows to simplify this expression to
\begin{align*}
Q_{BL} = \frac{N}{2\pi} \tan^{-1} \left[ \frac{8 \langle \hat{\bf S}_{1} \cdot [ \hat{\bf S}_{2} \times \hat{\bf S}_{3}]\rangle}{1 + 12\langle \hat{\bf S}_{1} \hat{\bf S}_{2}\rangle} \right].
\end{align*}
Expanding $\tan^{-1}(x)$ to first order in $x$ results in the following expression
\begin{align*}
Q_{BL} \simeq \alpha \frac{N}{\pi} \langle \hat{\bf S}_{1} \cdot [ \hat{\bf S}_{2} \times \hat{\bf S}_{3}]\rangle,
\end{align*}
where
$
\alpha = \frac{4}{1 + 12\langle \hat{\bf S}_{1} \hat{\bf S}_{2}\rangle}
$.
\begin{figure}[t!]
	\includegraphics[width=0.95\columnwidth]{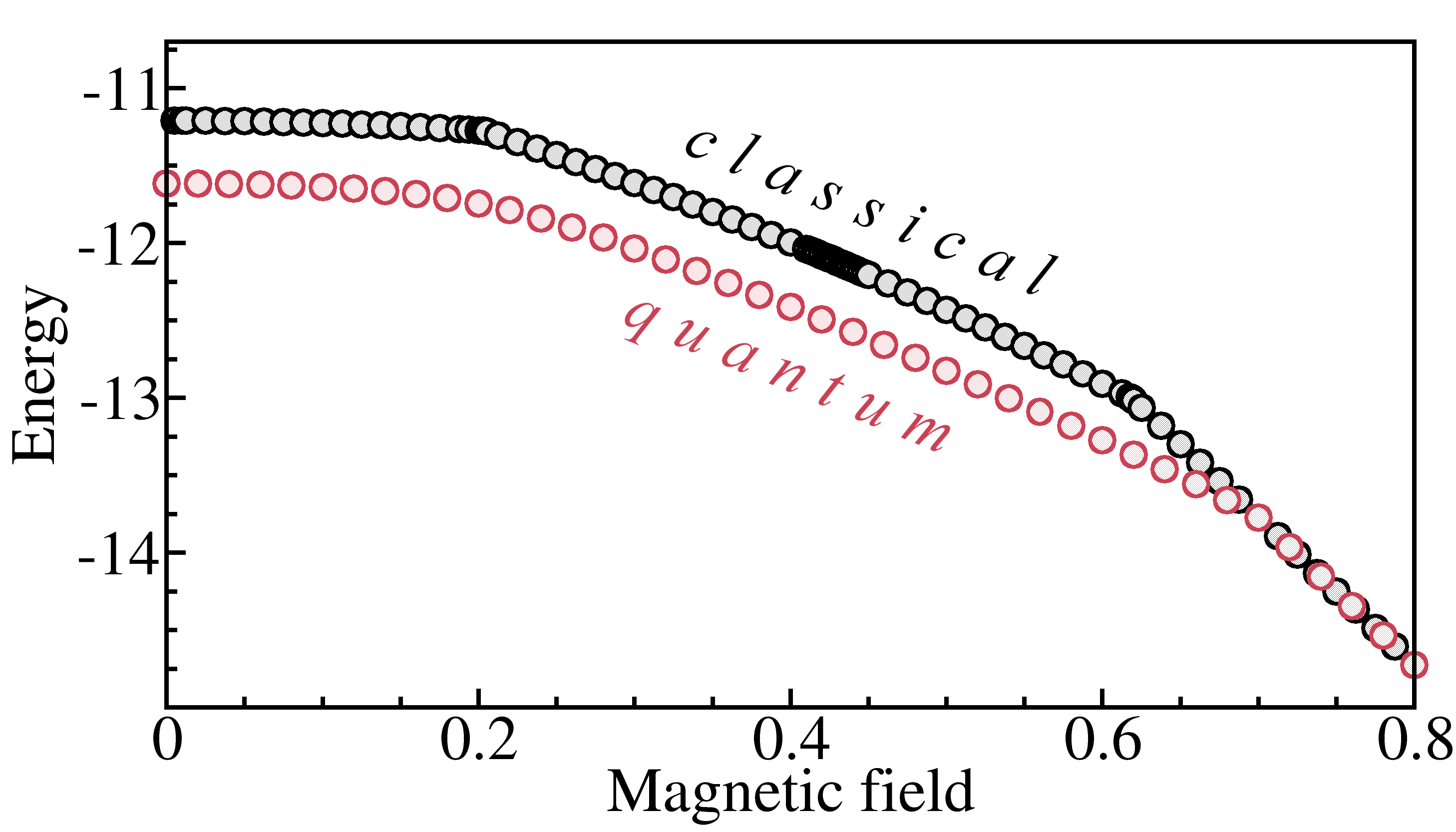}
	\caption{\label{fig:energies} Comparison of the classical (black circles) and quantum (red circles) energies of the ground state of the system. Parameters used for the calculation are the same as in Fig.~2 of the main text.}
\end{figure}
When the characteristic size of a skyrmion is much larger than the lattice constant, the magnetization varies slowly in space. Then, the nearest-neighbour spins are approximately aligned, and we get a short-range ferromagnetic order. Then, the spin-spin correlation function $\langle \hat{\bf S}_{1} \hat{\bf S}_{2}\rangle \simeq \langle \hat{\bf S}_{1}\rangle \, \langle \hat{\bf S}_{2}\rangle = 1/4$, leading to $\alpha = 1$. The three-spin correlation function can also be decoupled as
\begin{align*}
Q_{BL} &\simeq \frac{N}{\pi}\langle \hat{\bf S}_{1} \cdot [ \hat{\bf S}_{2} \times \hat{\bf S}_{3}]\rangle  
\simeq \frac{1}{\pi} \sum_{\langle ijk \rangle} \langle \hat{\bf S}_{i}\rangle \cdot [ \langle \hat{\bf S}_{j}\rangle \times \langle \hat{\bf S}_{k}\rangle ] \notag\\
&= \frac{1}{8\pi} \sum_{\langle ijk \rangle} {\bf m}_{i} \cdot [ {\bf m}_{j} \times {\bf m}_{k}],
\end{align*}
which results in the discrete analog of the standard formula for a classical skyrmion number.

\begin{figure*}[t!]
	\includegraphics[width=1\linewidth]{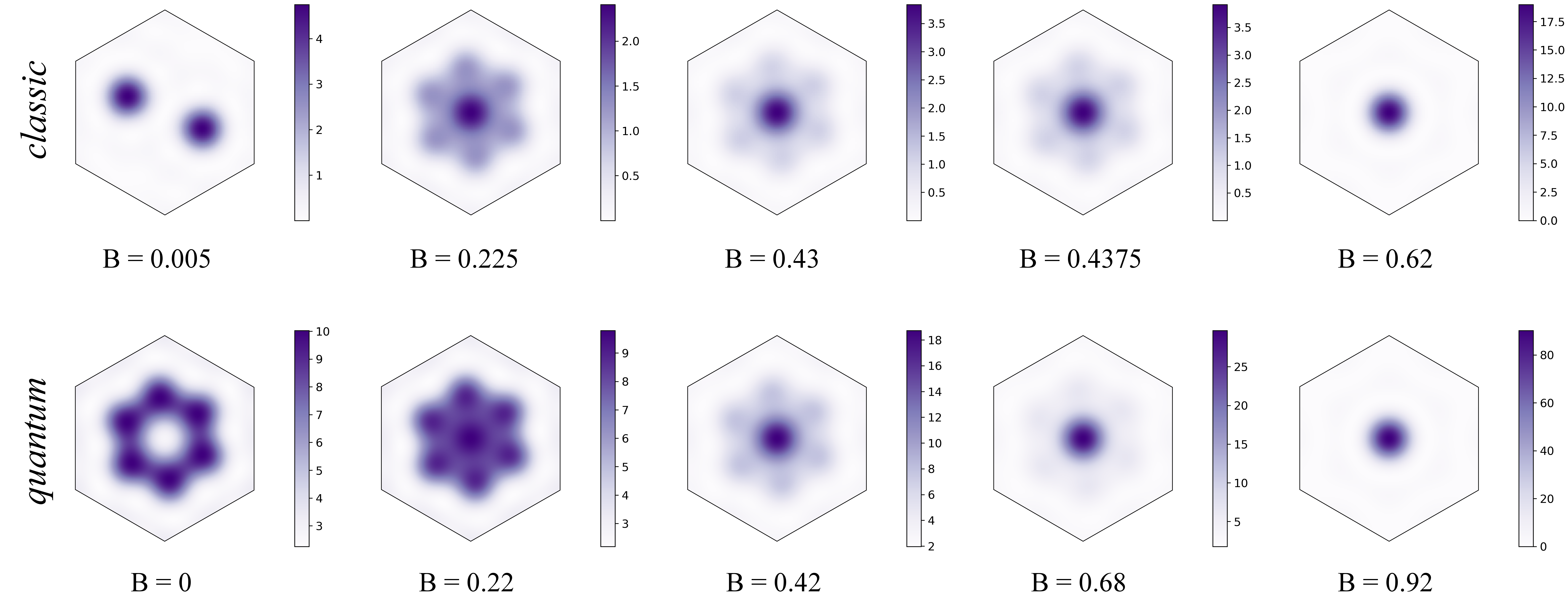}
	\caption{\label{Sfig:struct_factors} Momentum-space representation of the longitudinal spin structural factors for the classical (top row) and quantum (bottom row) system for different values of the magnetic field. Color bars indicate the intensity of Bragg peaks for each panel.}
\end{figure*}

We would like to note that the quantum skyrmion state has a lower energy than the corresponding classical skyrmion solution due to quantum fluctuations, as demonstrated in Fig.~\ref{fig:energies}. 
Thus, this novel quantum state represents the true ground state of the system. 

\section*{Two-spin correlation functions}

\begin{figure}[t!]
	\includegraphics[width=1\columnwidth]{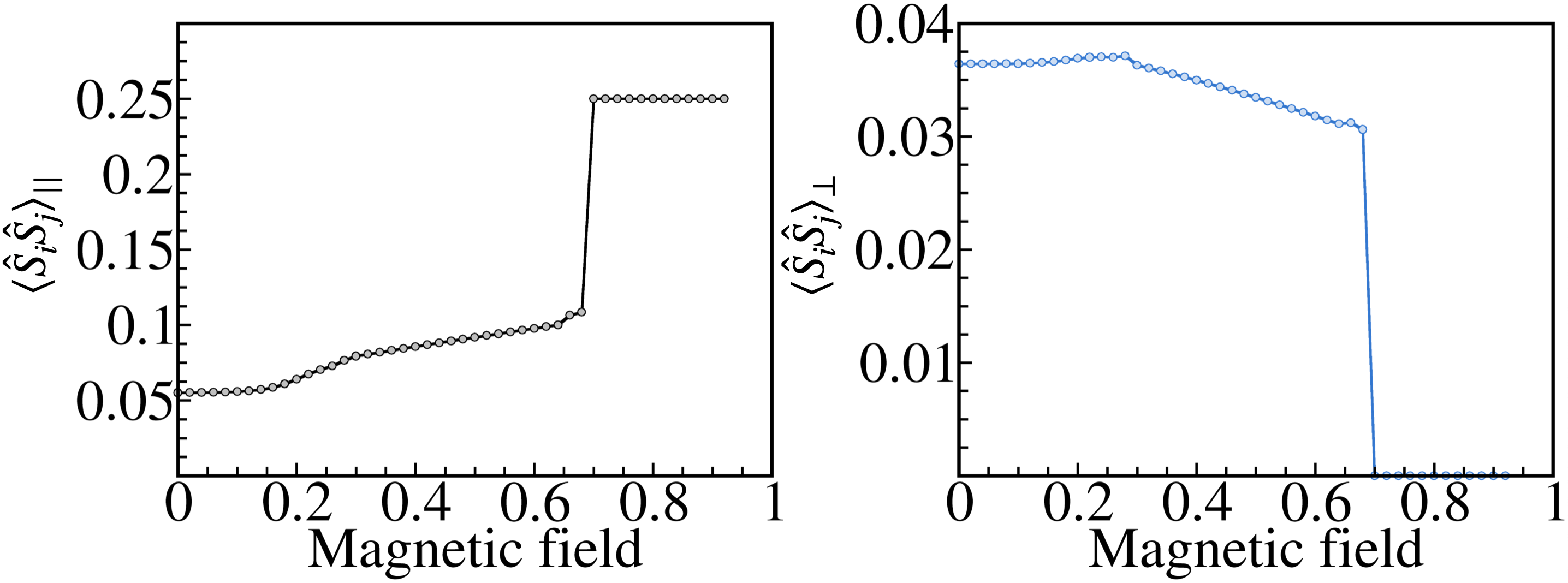}
	\caption{\label{Sfig:real-2spins} Longitudinal (left) and transverse (right) real-space spin-spin correlation function between two nearest-neighbor spins as a function of the magnetic field.}
\end{figure}

Spin-spin correlation function (structural factor) provides an important information on the magnetic excitations in a physical system.
Momentum-space representation of the longitudinal spin structural factors ${\langle \hat S_{\bf q} \hat S_{-{\bf q}} \rangle_{\parallel}}$ is presented in Fig.~\ref{Sfig:struct_factors} for different values of the magnetic field $B$. The detailed analysis of the position of Bragg peaks is presented in the main text.
Fig.~\ref{Sfig:real-2spins} shows the magnetic field dependence of the longitudinal ${\langle \hat S_{i} \hat S_{j} \rangle_{\parallel}}$ and transverse ${\langle \hat S_{i} \hat S_{j} \rangle_{\bot}}$ real-space spin-spin correlation function between neighboring $\langle ij \rangle$ spins. 
Here, ${\langle \hat S \hat S \rangle_{\parallel} = \langle \hat S^{z} \hat S^{z} \rangle}$ and ${\langle \hat S \hat S \rangle_{\bot} = \langle \hat S^{x} \hat S^{x} \rangle + \langle \hat S^{y} \hat S^{y} \rangle}$. 
Remarkably, we find that the structural factor is not constant in the skyrmionic phase.

\begin{figure}[t!]
	\includegraphics[width=1\columnwidth]{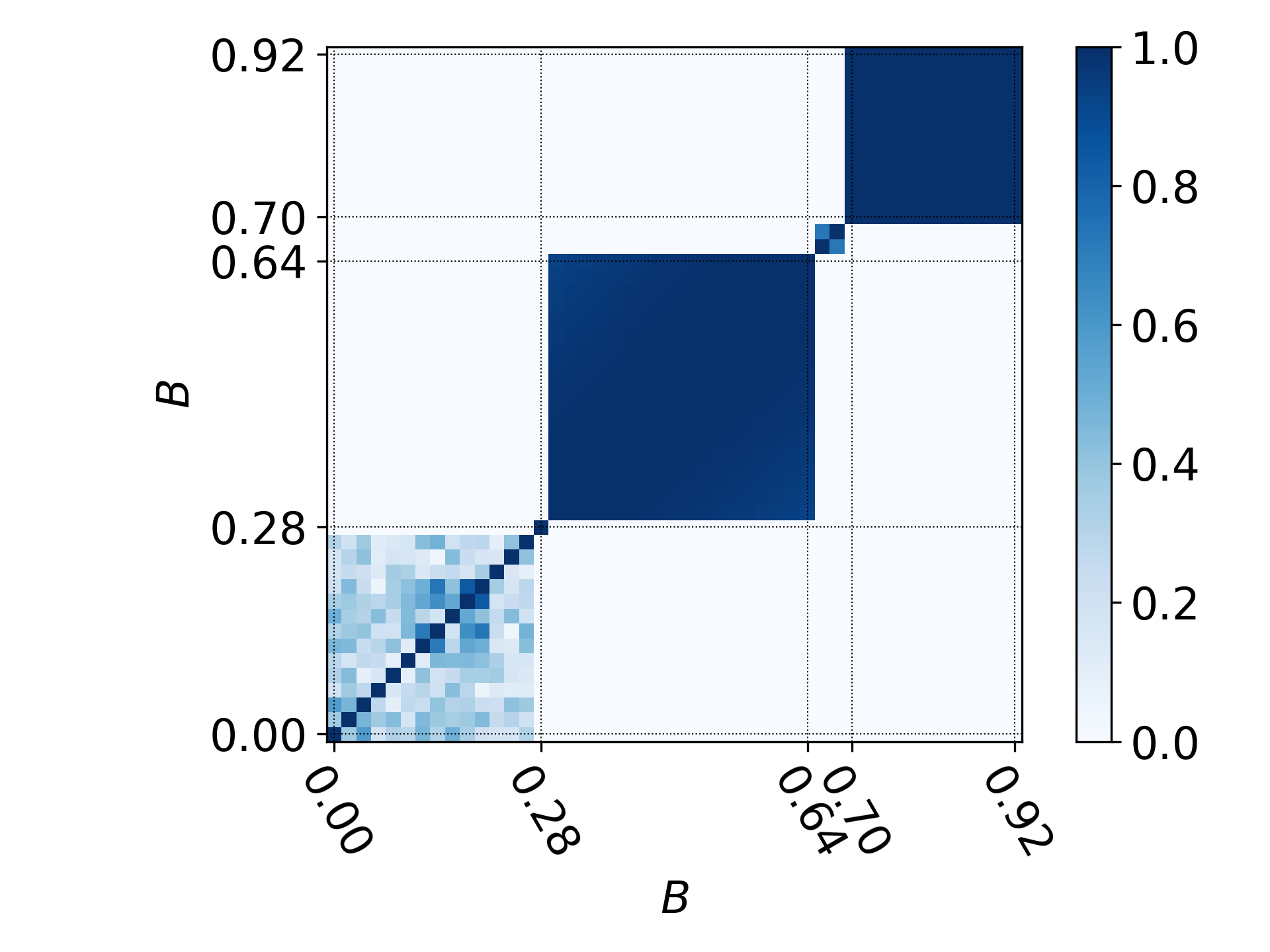}
	\caption{\label{Sfig:fidelity} Fidelity of the ground states calculated for the quantum problem.}
\end{figure}

\section*{How to distinguish quantum skyrmions of different types}

Another useful quantity to supplement the scalar chirality is the fidelity~\cite{Jozsa} $F_{\alpha\beta} = |\braket{\Psi_\alpha| \Psi_\beta}|$, which measures the overlap between two quantum states $\ket{\Psi_\alpha}$ and $\ket{\Psi_\beta}$. This quantity is shown in Fig.~\ref{Sfig:fidelity} as a function of the two magnetic fields at which the ground state eigenfunctions $ \ket{\Psi_{\alpha,\beta}} = \ket{\Psi(B_{\alpha,\beta})}$ were calculated. The intensity of the plot corresponds to the value of the fidelity $F_{\alpha\beta}$. Here, we observe four different phases that can be clearly isolated by exactly zero fidelity between their ground states. It is again truly remarkable that the fidelity approaches unity ($F_{\alpha\beta}>0.92$) in the intermediate field regime, which shows that for this range of magnetic fields the quantum ground state of the system indeed represents a unique quantum skyrmion state. Moreover, the fidelity inside the skyrmion phase is almost as uniform as in the fully polarized one, where it is identically equal to 1 due to the trivial product structure of the ground state of the system. By contrast, the low-field phase is more sensitive to a change of magnetic field: the value of the fidelity is unity only for equal values of the magnetic field $B_{\alpha}=B_{\beta}$. A similar behaviour is observed in a narrow window of magnetic fields $ 0.66 \le B \le 0.68 $ at the end of the intermediate field phase right before the transition to a polarized state. This result is consistent with the behavior of the quantum chirality presented in Fig.~2\,(B) of the main text.

It is important to note that quantum skyrmions stabilized at the same magnetic field but with differently oriented vector of the Dzyaloshinskii-Moriya interaction
can only be distinguished by the fidelity. For instance, one can obtain quantum skyrmions of Bloch or N\'eel types taking the in-plane anisotropic exchange interaction to be parallel or orthogonal to the real-space vector that connects neighboring lattice sites, respectively. In the classical case, the specific skyrmion type can be recognized by the magnetization density distribution. By contrast, the quantum Bloch and N\'eel skyrmions will be indistinguishable from the point of view of the magnetization, spin structural factors and quantum scalar chirality. The only quantity that can distinguish between these two quantum states is the fidelity, since $|\braket{\Psi_\text{N\'eel}| \Psi_\text{Bloch}}| \simeq 0$. 
However, the case when two different types of skyrmions appear in the same magnetic material is rather special.

\end{document}